\documentclass[a4paper,USenglish]{lipics-v2019}

\usepackage{amssymb, amsmath}
\usepackage{graphicx}
\usepackage{nicefrac}
\usepackage{sidecap}
\usepackage{url}
\usepackage{color}
\usepackage{rotating}
\usepackage{enumerate}
\usepackage{sidecap}
\usepackage{xspace}

\usepackage{thm-restate}
\usepackage[disable]{todonotes}

\let\epsilon\varepsilon

\def\qed{\rule{1.5mm}{1.5mm}\\ }
\newcommand{\ignore}[1]{}
\newcommand{\pbs}{\textsc{PBS}\xspace}
\newcommand{\starcover}{\textsc{StCo}\xspace}

\newcommand{\minBends}{\textsc{Min-Bends}\xspace}
\newcommand{\minSegments}{\textsc{Min-Segments}\xspace}

\newcommand{\mids}{\textsc{MIDS}\xspace}
\newcommand{\minimumMaximalIndependentSet}{\textsc{Minimum-Maximal-Independent-Set}\xspace}
\newcommand{\sat}{\textsc{SAT}\xspace}
\newcommand{\threeSAT}{\textsc{3-SAT}\xspace}
\newcommand{\p}{$\mathit{P}$\xspace}
\newcommand{\np}{$\mathit{NP}$\xspace}
\newcommand{\Oh}{\mathcal{O}}

\graphicspath{{figures/}}

\newenvironment{myproof}[1][\proofname]{\par
	\normalfont 
	\trivlist
	\item[\hskip\labelsep
	\color{darkgray}\sffamily\bfseries
	#1.]\ignorespaces
}{%
	\hfill\qedsymbol\endtrivlist
}

\title{Simplification of Polyline Bundles}

\titlerunning{Simplification of Polyline Bundles}

\author{Joachim Spoerhase}{Aalto University, Finland\\ University of W\"urzburg, Germany}{joachim.spoerhase@aalto.fi}{https://orcid.org/0000-0002-2601-6452}{Funded by European Research Council (ERC) under the European Union's Horizon 2020 research and innovation programme (grant agreement No. 759557).}
\author{Sabine Storandt}{University of Konstanz, Germany}{sabine.storandt@uni-konstanz.de}{}{Funded by the Deutsche Forschungsgemeinschaft (DFG, German Research Foundation) – Project-ID 50974019 – TRR 161.}
\author{Johannes Zink }{University of W\"urzburg, Germany}{zink@informatik.uni-wuerzburg.de}{https://orcid.org/0000-0002-7398-718X}{}

\authorrunning{J.~Spoerhase, S.~Storandt, and J.~Zink}

\Copyright{Joachim Spoerhase, Sabine Storandt, and Johannes Zink}

\ccsdesc[500]{Theory of computation~Approximation algorithms analysis}
\ccsdesc[500]{Theory of computation~Computational geometry}
\ccsdesc[300]{Theory of computation~Problems, reductions and completeness}
\ccsdesc[300]{Theory of computation~Fixed parameter tractability}

\keywords{Polyline Simplification, Bi-criteria Approximation, Hardness of Approximation, Geometric Set Cover}

\nolinenumbers 

\hideLIPIcs  

\begin{document}

\maketitle

\begin{abstract}
	We propose and study a generalization to the well-known problem of polyline simplification. Instead of a single polyline, we are given a set of $\ell$ polylines possibly sharing some line segments and bend points. Our goal is to minimize the number of bend points in the simplified bundle with respect to some error tolerance $\delta$ (measuring Fr\'echet distance) but under the additional constraint that shared parts have to be simplified consistently.
	We show that polyline bundle simplification is \np-hard to approximate within a factor~$n^{\frac{1}{3} - \varepsilon}$
	for any $\varepsilon > 0$ where $n$ is the number of bend points in the polyline bundle. This inapproximability even  applies to instances with only $\ell=2$ polylines. However, we identify the sensitivity of the solution to the choice of $\delta$ as a reason for this strong inapproximability. In particular, we prove that if we allow $\delta$ to be exceeded by a factor of $2$ in our solution, we can find a simplified polyline bundle with no more than $\Oh(\log (\ell + n)) \cdot OPT$ bend points in polytime, providing us with an efficient bi-criteria approximation.
	As a further result, we show fixed-parameter tractability in the number of shared bend points.
\end{abstract}

\section{Introduction}
Visualization of geographical information is a task of high practical relevance, e.g., for the creation of online maps.
Such maps are most helpful if the information is neatly displayed and can be grasped quickly and unambiguously.
This means that the full data often needs to be filtered and abstracted.
Many important elements in maps like borders, streets, rivers, or trajectories are displayed as polylines (also known as polygonal chains).
For such a polyline,  a simplification is supposed to be as sparse as possible and as close to the original as necessary.

A simplified polyline is usually constructed by a subset of bend points of the original polyline such that the (local) distance to the original polyline does not exceed a specifiable value according to a given distance measure, e.g., Fr\'echet distance 
or the Hausdorff distance. 
The first such algorithm, which is still of high practical importance, was proposed by Ramer~\cite{Ramer1972} and by Douglas and Peucker~\cite{Douglas1973}.
Hershberger and Snoeyink~\cite{Hershberger1992} proposed an implementation of this algorithm that runs in $\Oh(n \log n)$ time, where $n$ is the number of bend points in the polyline.
It is a heuristic algorithm as it does not guarantee optimality (or something close to it) in terms of retained bend points.
An optimal algorithm in this sense was first proposed by Imai and Iri~\cite{Imai1988}.
Chan and Chin~\cite{Chan1996} improved the running time of this algorithm to~$\Oh(n^2)$ for the Hausdorff distance.
For the Fr\'echet distance, the optimal solution can be determined in time~$\Oh(n^3)$ as described by Godau~\cite{Godau91}.

We remark that all of these algorithms consider the distance \emph{segment-wise}.
This is, the distance between each segment of the simplification and its corresponding sub-polyline of the input polyline does not exceed the given threshold.
We adhere to this widespread approach.
Intuitively and from an application point of view, it makes sense to map a point~$p$ of the input polyline only to a point of a segment of the simplification ``spanning over''~$p$ with respect to the input polyline as this ensures a certain degree of locality.
However, the general unrestricted approach has also received attention in the literature.
Here, the Hausdorff or Fr\'echet distance between the input polyline and the simplification as a whole polyline is considered.
For the (undirected) Hausdorff distance, this problem becomes \np-hard~\cite{Kreveld2020}
and for the Fr\'echet distance, there is an $\Oh(kn^5)$ time algorithm, where $k$ is the output complexity of the simplification~\cite{Kreveld2020}.
The problem variant where in addition the  requirement is dropped that all bend points of the simplification must be bend points of the input polyline, is called a \emph{weak} simplification.
Agarwal et al.~\cite{Agarwal2005} show that an optimal simplification under the segment-wise Fr\'echet distance with distance threshold~$\delta$, as computable using the algorithm by Imai and Iri, has no more bend points than an optimal weak simplification with distance threshold~$\delta / 4$.
We note that computing the Fr\'echet distance between two polylines can be solved in polynomial time~\cite{Alt1995}, but may become \np-hard
when considering additional properties like allowing to take shortcuts, which replace outliers in one of the polylines~\cite{Buchin2014}.

\begin{figure}
	\centering
	\begin{minipage}{0.48 \textwidth}
		\centering
		\includegraphics[page=1, scale=1.4]{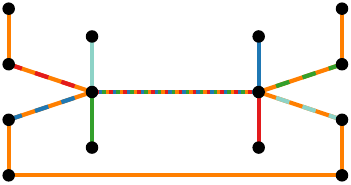}
	\end{minipage}
	\hfill
	\begin{minipage}{0.48 \textwidth}
		\centering
		\includegraphics[page=2, scale=1.4]{increasing-complexity}
	\end{minipage}
	
	\caption{Example where the total complexity increases if each polyline is simplified independently. Left: Initial bundle of polylines. Right: Bundle of independently simplified polylines.}
	\label{fig:increasing-complexity}
\end{figure}%
\begin{figure}
	\centering
	\begin{minipage}{0.48 \textwidth}
		\centering
		\includegraphics[page=1, scale=1.15]{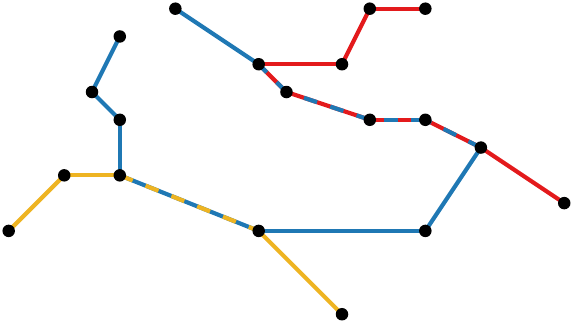}
	\end{minipage}
	\hfill
	\begin{minipage}{0.48 \textwidth}
		\centering
		\includegraphics[page=2, scale=1.15]{simplification-example}
	\end{minipage}
	\caption{Example of a bundle of three polylines before and after consistent simplification.}
	\label{fig:simplification-example}
\end{figure}

\paragraph*{From a Single Polyline to a Bundle of Polylines}
On a map, there are usually multiple polylines to display.
Such polylines may share bend points (bends) and line segments between bends (segments) sectionwise.
We call them a \emph{bundle} of polylines.
One example is a schematic map of a public transport network where 
bus lines are the polylines and these share some of the stations and legs.

One might consider simplifying the polylines of a bundle independently.
This has some drawbacks, though.
On the one hand, the total complexity might even increase when the shared parts are simplified in different ways; see Figure~\ref{fig:increasing-complexity}.
On the other hand, it might suggest a misleading picture when we remove common segments and bends of some polylines, but not of all.
Therefore, we require that a bend in a simplification of a bundle of polylines is either kept in all polylines containing it or discarded in all polylines. Our goal is then to minimize the total number of bend points that have to be kept.
In Figure~\ref{fig:simplification-example}, we give an example of a simplification of a bundle of polylines.


\paragraph*{Related Work}
Polyline bundles were studied before in different contexts. In~\cite{buchin2018group}, the goal is to interfere a concise graph which represents all trajectories in a given bundle sufficiently well. But this approach primarily aims at retrieving split and merge points of trajectories correctly and does not produce a simplification of each trajectory in the bundle. Methods for map generation  based on movement trajectories~\cite{he2018roadrunner} 
have a similar scope but explicitly allow to discard outliers and to unify sufficiently similar trajectories, which is not allowed in our setting.

Agarwal et al.~\cite{Agarwal2005} describe an $\Oh(n \log n)$ time approximation algorithm for (classical) polyline simplification under the Fr\'echet distance.
It is an approximation algorithm in the sense that the output simplification for distance threshold~$\delta$ has at most as many bends as an optimal solution with distance threshold~$\delta / 2$.
In Theorem~\ref{thm:min-bends-hard-to-approx}, we also relate the size of our approximate solution respecting a distance thershold of~$\delta$ to an optimal solution with distance threshold~$\delta / 2$.

There is also a multitude of polyline simplification problem variants for single polylines which involve additional constraints. One important variant is the computation of the smallest possible simplification of a single polyline which avoids self-intersection~\cite{Berg1998}. Another practically relevant variant is the consideration of topological constraints. For example, if the polyline represents a country border, important cities within the country should remain on the same side of the polyline after simplification. It was proven that those problem variants are hard to approximate within a factor $n^{\frac{1}{5}-\varepsilon}$ \cite{estkowski2001simplifying}. Hence, in practice, they are typically tackled with heuristic approaches~\cite{estkowski2001simplifying,funke2017map}.

Note that the only allowed inputs to  those problem variants are either a single polyline without self-intersections or a set of polylines without self-intersections and without common bends or segments (except for common start or end points). In contrast, we explicitly allow non-planar inputs and polyline bundles in which bends and segments may be shared among multiple polylines. We also remark that the known results on hardness of approximation of these problems heavily rely on the constraint that feasible solutions are still non-intersecting. Since we do not require this, we have to resort to different techniques.

\paragraph*{Contribution}
We introduce the optimization problem of polyline bundle simplification, where we are given $\ell$ polylines on an underlying set of $n$ points as well as an error bound $\delta$ and seek to find a simplified polyline bundle with the smallest possible number of remaining points, where each simplified polyline has a Fr\'echet distance of no more than $\delta$ to the original polyline and the simplification is consistent for shared parts.

While the optimal simplification of a single polyline can be computed in polynomial time, we show that polyline bundle simplification is \np-hard to approximate within a factor~$n^{\frac{1}{3} - \varepsilon}$ for any $\varepsilon > 0$. This result applies already to bundles of two polylines, hence excluding an efficient FPT-algorithm depending on parameter~$\ell$.

On the positive side, we show that this strong inapproximability can be overcome when relaxing the error bound $\delta$ slightly. In particular, we design an efficient bi-criteria approximation algorithm. 
Here, we allow the simplified polylines in our solution to have a Fr\'echet distance of $2\delta$ instead of only $\delta$ to the original polylines. We can then approximate the optimal solution for the original choice of $\delta$ within a factor logarithmic in the input size.
As the choice of $\delta$ for real-world problems often is made in a rather ad hoc fashion and uncertainties with respect to the precision of the input polylines have to be factored in as well, we deem our bi-criteria approximation to be of high practical relevance.

We furthermore show that, while the number of polylines in the bundles is not suitable to obtain an FPT-algorithm, the problem of polyline bundle simplification is indeed fixed-parameter tractable in the number of bend points that are shared among the polylines.
%

\section{Formal Problem Definition}
\label{sec:problem-statement}

An instance of the \emph{polyline bundle simplification} problem  (from now on abbreviated by \pbs) is specified by a triple $(B, \mathcal{L}, \delta)$ , where $B = \{b_1, \dots, b_n\}$ is a set of $n$ points (\emph{bends}) in the plane, a \emph{polyline bundle} $\mathcal{L}$, which is a set $\mathcal{L} = \{L_1, \dots, L_\ell\}$  of $\ell$~polylines $L_i=(s_i, \dots, t_i)$ represented as lists of points from~$B$, and a distance parameter~$\delta$,
which specifies a threshold for the the maximum (segment-wise) Fr\'echet distance between original and simplified polyline bundle.
Each polyline $L_i$ ($i \in \{1, \dots, \ell\}$) is simple in the sense that each bend of $B$ appears at most once in its list.
	
\begin{definition}[Polyline Bundle Simplification]
	Given a triple $(B, \mathcal{L}, \delta)$, the goal is to obtain a minimum size subset $B^* \subseteq B$ of points, such that for each polyline $L_i \in \mathcal{L}$ its induced simplification $S_i$ (which is $L_i \cap B^*$ while preserving the order of points)
	\begin{itemize}
		\item contains the start and the end point of $L_i$, i.e., $s_i, t_i \in S_i$, and
		\item has a segment-wise Fr\'echet distance of at most $\delta$ to $L_i$, i.e., for each line segment $(a,b)$ of $S_i$ and the corresponding sub-polyline of $L_i$ from $a$ to $b$, abbreviated by $L_i[a,\dots,b]$, we have $d_\textnormal{Fr\'echet}((a,b), L_i[a,\dots,b]) \leq \delta$.
	\end{itemize}
\end{definition}

For the sake of self-containedness we restate the definition of the Fr\'echet distance below.
\begin{definition}[Fr\'echet Distance]
	Between two polylines $L_1 = (b_{1,1}, b_{1, 2}, \dots, b_{1, |L_1|})$ and $L_2 = (b_{2,1}, b_{2, 2}, \dots, b_{2, |L_2|})$ in the Euclidean plane, the \emph{Fr\'echet distance}~$d_\textnormal{Fr\'echet}(L_1, L_2)$ is
	\begin{equation*}
		d_\textnormal{Fr\'echet}(L_1, L_2) := \inf_{\alpha, \beta} \max_{t \in [0, 1]} \| c_{L_1}(\alpha(t)) - c_{L_2}(\beta(t)) \| \, ,
	\end{equation*}
	where $\alpha \colon [0, 1] \rightarrow [1, |L_1|]$ and $\beta \colon [0, 1] \rightarrow [1, |L_2|]$ are continuous and non-decreasing functions with $\alpha(0) = \beta(0) = 1$, $\alpha(1) = |L_1|$, $\beta(1) = |L_2|$, \\
	and $c_{L_i} \colon [1, |L_i|] \rightarrow \mathbb{R}^2$ with $c_{L_i} \colon x \mapsto (\lfloor x \rfloor + 1 - x) b_{i, \lfloor x \rfloor} + (x - \lfloor x \rfloor) b_{i, \lceil x \rceil}$.
\end{definition}


\section{Hardness of Polyline Bundle Simplification}
\label{sec:reductionFromMIDS}


In this section, we describe a polynomial-time reduction from Minimum Independent Dominating Set (\mids) to \pbs to show \np-hardness and hardness of approximation.
In the \mids problem, we are given a graph $G = (V, E)$, where $V$ is the vertex set and $E$ is the edge set of~$G$.
We define $\hat{n} = |V|$ and $\hat{m} = |E|$.
The goal is to find a set $V^* \subseteq V$ of minimum cardinality that is a dominating set of~$G$ as well as an independent set in~$G$.
A dominating set contains for each vertex~$v$, $v$ itself or at least one of $v$'s neighbors.
An independent set contains for each edge at most one of its endpoints.
Halld{\'{o}}rsson~\cite{Halldorsson1993} has shown that \mids, which is also referred to as \minimumMaximalIndependentSet,
is \np-hard to approximate within a factor of~$|V|^{1 - \varepsilon}$ for any $\varepsilon > 0$.
In his proof, he uses a reduction from \sat to \mids: from a \sat formula~$\Phi$, he constructs a graph such that an algorithm approximating \mids would decide if~$\Phi$ is satisfiable.
We observe that this reduction is still correct if $\Phi$ is a \threeSAT formula.
Moreover, we observe that the number of edges in the graph constructed in this reduction by a \threeSAT formula is linear in the number of vertices.
Thus, we conclude the following corollary and assume henceforth that we reduce only from sparse graph instances of \mids, in other words, $\hat{m} \leq c \hat{n}$ for some sufficiently large constant~$c$.
\begin{corollary}
	\label{cor:mids-sparse-graphs}
	\mids on graphs of $\hat{n}$ vertices and $\Oh(\hat{n})$ edges, i.e., sparse graphs, is \np-hard to approximate within a factor of~$\hat{n}^{1 - \varepsilon}$ for any $\varepsilon > 0$.
\end{corollary}

\begin{figure}[t]
	\centering
	\begin{subfigure}[b]{0.26 \linewidth}
		\centering
		\includegraphics[page=1,scale=1.1,trim=235 147 529 0,clip]{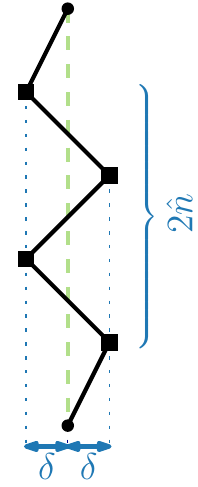}
		\caption{Vertex gadget; bends that may be shared with edge or neigh\-borhood gad\-gets are drawn as squares}
		\label{fig:MIDS-reduction-vertex-gadget}
	\end{subfigure}
	\hfill
	\begin{subfigure}[b]{0.7 \linewidth}
			\centering
			\includegraphics[page=3,scale=1.24,trim=226 177 372 34,clip]{mids-reduction}
			\caption{Edge gadget for an edge $u v$; the second and second last bend (drawn as squares) are shared with the vertex gadgets of $u$ and $v$, respectively. If and only if at least one of the two shared bends is skipped, we can skip all $2\hat{n}^2 + 1$ inner bends.}
			\label{fig:MIDS-reduction-edge-gadget}
	\end{subfigure}
		
	\medskip
	
	\begin{subfigure}[b]{1 \linewidth}
			\centering
			\includegraphics[page=11,scale=1.4,trim=264 0 276 205,clip]{mids-reduction}
			\caption{Neighborhood gadget for a vertex $v$; the bends drawn as squares are shared with the vertex gadgets of~$v$ and $v$'s neighbors in the graph. Only if we keep at least one of the shared bends, we can skip almost all bends of the gadget.}
			\label{fig:MIDS-reduction-neighborhood-gadget}
	\end{subfigure}

	\medskip
	
	\begin{subfigure}[t]{1 \linewidth}
		\centering
		\includegraphics[page=10,scale=0.48,trim=0 106 0 0,clip]{mids-reduction}
		\caption{Combination of three vertex gadgets (for the vertices $v_1, v_2, v_3$) with two edge gadgets (for the edges $v_1 v_2$ and $v_2 v_3$) and a neighborhood gadget for the vertex $v_2$.}
		\label{fig:MIDS-reduction-combined-example}
	\end{subfigure}
	\caption{Schematization of our reduction from \mids to \pbs. Shortcuts are indicated by dashed green line segments. Dashed red line segments between two bends indicate that there is no shortcut.
	The vertices in our minimum independent dominating set are precisely the ones for which we do not take the shortcut of the corresponding vertex gadgets.
	}
	\label{fig:MIDS-reduction}
\end{figure}

In our reduction, we use three types of gadgets, which are in principle all lengthy zigzag pieces.
We use
	\emph{vertex gadgets} to indicate whether a vertex is in the set $V^*$ or not,
	\emph{edge gadgets} to enforce the independent set property, and
	\emph{neighborhood gadgets} to enforce the dominating set property.
See Fig.~\ref{fig:MIDS-reduction} for an overview.
We define our gadgets in terms of an arbitrary~$\delta$ (threshold for the maximum Fr\'echet distance) and some $\gamma \leq 2 \delta / (10 \hat{n}^2 + 5)$.
Note that our problem setting allows overlaps of different polylines without having a common bend or segment (non-planar input).
In our reduction there can also be overlaps, which do not affect the involved polylines locally.
\bigskip\\\textbf{Vertex Gadget.}
For each vertex, we construct a \emph{vertex gadget} (see Figure~\ref{fig:MIDS-reduction-vertex-gadget}), which we arrange vertically next to each other on a horizontal line in arbitrary order and with some distance $x_\textrm{spacing} \ge (2\hat{n}^2+2) 3\delta$ between one and the next vertex gadget.

A vertex gadget has $2 \hat{n} + 2$ bends arranged in a zigzag course with x-distance $2 \delta$ ($\delta$ for the first and the last segment) and y-distance $3 \delta$ between each two consecutive bends.
\begin{claim}
	In a vertex gadget, there is precisely one shortcut, which starts at the first and ends at the last bend.
\end{claim}
Clearly, the line segment from the first to the last bend has Fr\'echet distance at most $\delta$ to the other bends and segments of the vertex gadget. 
Moreover, observe that there is no shortcut starting or ending at any inner bend.
Thus, either none or all inner bends are skipped.
We say that the corresponding vertex is in~$V^*$ if and only if we do not skip the inner vertices.
\bigskip\\\textbf{Edge Gadget.}
For each edge $\{u,v\}$, we construct an \emph{edge gadget} (see Figure~\ref{fig:MIDS-reduction-edge-gadget}) being a zigzag course with $2\hat{n}^2 + 5$ bends and sharing its second and second last bend with one of the two corresponding vertex gadgets---the vertex gadgets of $u$ and~$v$.
All neighboring bends from the second to the second last are equidistant in x-dimension, while the first and second bend, and the second last and last bend have the same x-coordinate.
In y-dimension, the first and the last bend are $2/5 \delta + \gamma$ below the second and second last bend, respectively.
The other bends are $3/5 \delta - \gamma$ above the second bend or $3/5 \delta$ below the first bend.

\begin{restatable}[]{claim}{claimEdgegadget}
	\label{clm:claimEdgegadget}
	In an edge gadget, there are precisely three long shortcuts.
	These are (i) from the first to the last bend, (ii) from the first to the second last bend, and (iii) from the second to the last bend.
	Beside these three shortcuts, there are $\leq 4$ more shortcuts, which skip only the second and the second last bend (and possibly also the third and third last bend).
	There is no shortcut not skipping one of the shared bends, i.e., the second or the second last bend.
\end{restatable}
\noindent In Appendix~\ref{app:reductionFromMIDS}, we argue that Claim~\ref{clm:claimEdgegadget} is correct.
It follows that not skipping one of the two shared bends is a relatively expensive choice in terms of retained bends.
Remember that not skipping one of the shared bends means not taking the shortcut in the corresponding vertex gadget, which means putting the corresponding vertex into~$V^*$.
So, skipping almost all bends in the edge gadget of~$\{u,v\}$ implies not having $u$ or $v$ in~$V^*$, which means respecting the independent set property for the edge $\{u,v\}$.
\bigskip\\\textbf{Neighborhood Gadget.}
For each vertex $v$, we construct a \emph{neighborhood gadget} (see Figure~\ref{fig:MIDS-reduction-neighborhood-gadget}).
This gadget shares a bend with every vertex gadget corresponding to a vertex of $\textrm{Adj}(v)$, which is $v$ and the vertices being adjacent to $v$.
These shared bends are on the same height.
The vertex gadgets of $\textrm{Adj}(v)$ appear in some horizontal order in our construction.
Say the corresponding vertices in order are $u_1, \dots, u_{|\textrm{Adj}(v)|}$.
Let the shared bends with $u_1$ and $u_{|\textrm{Adj}(v)|}$ be $b_1$ and $b_{|\textrm{Adj}(v)|}$, respectively, and define $t$ as the distance between $b_1$ and $b_{|\textrm{Adj}(v)|}$.
We place the first bend (the starting point) of the neighborhood gadget $4/5 \delta$ below and $3 t$ to the left of~$b_1$, where $t$ is the distance between $b_1$ and~$b_{|\textrm{Adj}(v)|}$, and let the second bend be $b_1$.
Symmetrically, we place the last bend (the end point) of the gadget $4/5 \delta$ below and $3 t$ to the right of~$b_{|\textrm{Adj}(v)|}$ and let the second last bend be $b_{|\textrm{Adj}(v)|}$.
Between each two bends $b_i$ and $b_{i + 1}$ shared with the vertex gadgets of $u_i$ and $u_{i + 1}$ for each $i \in \{1, \dots, |\textrm{Adj}(v)|-1\}$, we add a zigzag with $2 \hat{n}^2 + 1$ bends as in~Figure~\ref{fig:MIDS-reduction-neighborhood-gadget}.
\begin{restatable}[]{claim}{claimNeighborhoodgadget}
	\label{clm:claimNeighborhoodgadget}
	In a neighborhood gadget, the only shortcuts are (i) the shortcuts skipping only $b_i$ for $i \in \{1, \dots, |\textrm{Adj}(v)|\}$ and (ii) the shortcuts starting at the first bend or $b_i$ with $i \in \{1, \dots, |\textrm{Adj}(v)|\}$ and ending at the last bend or $b_j$ with $i < j \in \{1, \dots, |\textrm{Adj}(v)|\}$---except for the shortcut starting at the first and ending at the last bend.
\end{restatable}
\noindent In Appendix~\ref{app:reductionFromMIDS}, we argue that Claim~\ref{clm:claimNeighborhoodgadget} is correct.
Consequently, we can skip almost all bends in a neighborhood gadget if we keep at least one bend of $b_1, \dots, b_{|\textrm{Adj}(v)|}$.
If we skip all of them, we can skip no other bend. 
So, to avoid high costs, we must not take the shortcut of the vertex gadget of at least one vertex of~$\textrm{Adj}(v)$.
This means that we must, for each $v \in V$, add a vertex of $\textrm{Adj}(v)$ to $V^*$, which enforces the dominating set property.

Observe that all shared bends are shared between only two polylines---by a vertex gadget and either an edge gadget or a neighborhood gadget.
With $2 \hat{n}$ inner bends, a vertex gadget provides enough bends that are shared with the edge and neighborhood gadgets as a vertex is contained in at most~$\hat{n}$ neighborhoods and has at most $\hat{n}-1$ incident edges.
In the following lemma, we analyze the size of the constructed \pbs instance. 

\begin{restatable}[]{lemma}{midsReductionSize}
	\label{lem:mids-reduction-size}
	By our reduction, we obtain from an instance $G = (V, E)$ of \mids an~instance of \pbs with $n$ bends such that $n \leq 10 c \hat{n}^3$, where $\hat{n} = |V| \geq 2$, $|E| \leq c \hat{n}$ ($c \geq 1$ is~constant).
\end{restatable}

\begin{myproof}
	To count the bends of the vertex, edge, and neighborhood gadgets without double counting, we charge the shared bends to the vertex gadgets.
	All vertex gadgets together have
	$\hat{n} (2 \hat{n} + 2)$
	bends, all edge gadgets have
	$\hat{m} (2\hat{n}^2 + 3)$
	bends without shared bends, and all neighborhood gadgets have
	$2 \hat{m} \cdot  (2\hat{n}^2 + 1) + 2 \hat{n}$
	bends without shared bends.
	Summing these values up and using $\hat{m} = |E| \leq c \hat{n}$ yields (for $\hat{n} \ge 2$)
	\begin{equation}
	n = 2\hat{n}^2 + 2 \hat{n} + 2 \hat{m} \hat{n}^2 + 3 \hat{m} + 4 \hat{m} \hat{n}^2 + 2 \hat{m} + 2 \hat{n} \leq 6 c \hat{n}^3 + 2 \hat{n}^2 + (4 + 5 c) \hat{n} \leq 10 c \hat{n}^3 \, .
	\end{equation}
\end{myproof}
%
We say a simplification of an instance of \pbs obtained by this reduction \emph{corresponds} to an independent and dominating set~$V'$ and vice versa
if we take all ``long'' shortcuts in the vertex gadgets except for the ones corresponding to $V'$ and we skip all inner unshared bends in all edge and neighborhood gadgets,
which is possible since $V'$ is independent and dominating.
Observe that for each independent and dominating set there is precisely one corresponding simplification (which is also valid acc.\ to~$\delta$).

\begin{restatable}[]{lemma}{midsSolutionInPBS}
	\label{lem:mids-solution-in-pbs}
	Let $V'$ be a solution for an instance~$G = (V, E)$ of \mids.
	In the instance~$(B, \mathcal{L}, \delta)$ of \pbs obtained by our reduction, the size of the simplification corresponding to $V'$ is $2 \hat{n} (|V'| + c + 2)$, where $\hat{n} = |V|$ and $c \ge 1$ is constant.
\end{restatable}

\begin{myproof}
	Only for all $v \in V \setminus V'$, we take the shortcuts in the corresponding vertex gadgets in $(B, \mathcal{L}, \delta)$.
	This gives us
	$\left(\hat{n} - |V'|\right) \cdot 2 + |V'| \cdot (2 + 2 \hat{n}) = 2 \hat{n} \left(1 + |V'|\right)$
	remaining bends in all vertex gadgets combined.
	In the following, we will count shared bends for the vertex gadgets.
	We take a ``long'' shortcut in all of the edge gadgets.
	This gives us two remaining unshared bends in all edges gadgets ($c \hat{n} \cdot 2$ bends in total).
	Moreover, we skip all inner unshared bends in all of the neighborhood gadgets ($2 \hat{n}$ bends remaining).
	Altogether, this sums up to $2 \hat{n} (|V'| + 1 + c + 1)$.
\end{myproof}

By Lemma~\ref{lem:mids-solution-in-pbs}, we know that for an optimal solution~$V^*$ of an instance of~\mids, the corresponding simplification in the instance~$(B, \mathcal{L}, \delta)$ of~\pbs obtained by our reduction has size $2 \hat{n} (OPT_\mids + c + 2)$, where $OPT_\mids = |V^*|$ and which of course is at least the size~$OPT_\pbs$ of the optimal solution of $(B, \mathcal{L}, \delta)$.
We formalize this in the following corollary.

\begin{corollary}
	\label{cor:bound-of-opts-in-inapprox-reduction}
	For an instance~$G = (V, E)$ of \mids and the instance~$(B, \mathcal{L}, \delta)$ of \pbs obtained by our reduction from~$G$,
	$OPT_\pbs \le 2 \hat{n} (OPT_\mids + c + 2)$.
\end{corollary}

\begin{theorem}
	\label{thm:min-bends-hard-to-approx}
	\pbs is \np-hard to approximate within a factor of~$n^{\frac{1}{3} - \varepsilon}$ for any $\varepsilon > 0$,
	where $n$ is the number of bend points in the polyline bundle.
\end{theorem}

\begin{myproof}
	Assume that there is an approximation algorithm~$\mathcal{A}$ solving any instance of \pbs within a factor of $n^{\frac{1}{3} - \varepsilon}$ for some constant $\varepsilon > 0$ relative to the optimal solution.
	We can transform any instance~$G = (V, E)$ of \mids, where $\hat{n} = |V|, \hat{m} = |E|$, and $OPT_\mids = |V^*|$, this is the size of an optimal solution, to an instance~$(B, \mathcal{L}, \delta)$ of \pbs using the reduction described above in this section,
	where $|B| = n$ and the size of an optimal solution is~$OPT_\pbs$.
	
	Employing $\mathcal{A}$ to solve $(B, \mathcal{L}, \delta)$ yields a (simplified) polyline bundle~$\mathcal{L}_\mathcal{A}$.
	We denote the number of bends in $\mathcal{L}_\mathcal{A}$ by~$n_\mathcal{A}$ and we know that $n_\mathcal{A} \leq OPT_\pbs \cdot n^{\frac{1}{3} - \varepsilon}$ for some $\varepsilon > 0$.
	If all $(2\hat{n}^2 + 1)$-bend-sequences in all edge and neighborhood gadgets are skipped, we can immediately read an independent dominating vertex set $V' \subseteq V$ from the vertex gadgets where the shortcut is not taken.
	Otherwise, we replace $\mathcal{L}_\mathcal{A}$ such that it corresponds to any maximal independent set~$V' \subseteq V$ (which is always an independent and dominating set and can be found greedily in polynomial time).
	Observe that this can only lower the number of bends compared to a solution not skipping all $(2\hat{n}^2 + 1)$-bend-sequences in the edge and neighborhood~gadgets as in all vertex gadgets together we can skip at most $\hat{n} \cdot 2 \hat{n}$ bends.
	
	Using Lemma~\ref{lem:mids-solution-in-pbs} and Corollary~\ref{cor:bound-of-opts-in-inapprox-reduction}, we can state that
	\begin{align}
	\label{eq:n_AtoOPT_PBS}
		n^{\frac{1}{3} - \varepsilon} &\geq \frac{n_\mathcal{A}}{OPT_\pbs} \geq \frac{2 \hat{n} (|V'| + c + 2)}{2 \hat{n} (OPT_\mids + c + 2)} > \frac{|V'|}{OPT_\mids + c + 2} \, ,
	\end{align}
	which we can reformulate as
		$|V'| < n^{\frac{1}{3} - \varepsilon}(OPT_\mids + c + 2)$.
	We can assume that $OPT_\mids > c + 2$ as otherwise we could check all subsets of $V$ of size at most~$c + 2$ in polynomial time.
	Similarly, we can assume that $\hat{n}$ is large enough so that $\hat{n}^{2 \varepsilon} > 20 c$.
	Beside this, we apply Lemma~\ref{lem:mids-reduction-size} and obtain
	\begin{align}
		|V'| &< 2 n^{\frac{1}{3} - \varepsilon} OPT_\mids 
		\leq 2 \cdot (10c\hat{n}^3)^{\frac{1}{3} - \varepsilon} OPT_\mids\\
		&< 20 c \cdot \hat{n}^{1 - 3 \varepsilon} OPT_\mids
		< \hat{n}^{2 \varepsilon} \cdot \hat{n}^{1 - 3 \varepsilon} OPT_\mids
		= \hat{n}^{1 - \varepsilon} OPT_\mids \, .
		\label{eq:maintheorem}
	\end{align}
	
	Since we know that it is \np-hard to approximate \mids within a factor of $\hat{n}^{1 - \varepsilon}$ for any $\varepsilon > 0$, it follows that $\mathcal{A}$ cannot be a polynomial time algorithm, unless \p = \np.
	Or in other words, it is \np-hard to approximate \pbs within a factor of~$n^{\frac{1}{3} - \varepsilon}$ for any $\varepsilon > 0$.
\end{myproof}

%

Currently, we use one polyline per gadget.
So, our reduction uses $2\hat{n} + \hat{m}$ polylines.
We can reduce the number of polylines to two by connecting all vertex gadgets---one after the other---in arbitrary order by two segments, which gives us the first polyline, and by connecting all edge and neighborhood gadgets in arbitrary order by two segments, which gives us the second polyline.
The extra bend between each pair of new segments is placed far away from the construction, e.g. at $(\infty, \infty)$.
This never creates new shortcuts for skipping a bend in a vertex gadget or in a neighborhood gadget.
Yet, we might create new shortcuts that allow for additionally skipping the first and the last bend of an edge gadget.
However, we cannot skip any further bend unless the second or second last bend is skipped, which preserves the functionality of our gadget.
For the analysis, this gives us an additive constant of at most $2 \hat{n} + \hat{m}$ bends that cannot be skipped, which we can include to Inequalities~(\ref{eq:n_AtoOPT_PBS})--(\ref{eq:maintheorem}) in Theorem~\ref{thm:min-bends-hard-to-approx} with the same result to obtain the following~corollaries.
\begin{corollary}
	Even for instances of two polylines, \pbs is \np-hard to approximate within a factor of~$n^{\frac{1}{3} - \varepsilon}$
	for any $\varepsilon > 0$, where $n$ is the number of bend points in the polyline bundle.
\end{corollary}

\begin{corollary}
	\label{cor:pbsNotFPTinL}
	\pbs is not fixed-parameter tractable in the number of polylines~$\ell$.
\end{corollary}


\section{Bi-criteria Approximation for Polyline Bundle Simplification}
\label{sec:bi-criteria-approx}
In this section, we describe a bi-criteria approximation algorithm for \pbs. Conceptually, a bi-criteria approximation is a generalization of a (classical) approximation
where it is allowed to violate a certain constraint by a specific factor. In particular, an algorithm is called a bi-criteria $(\alpha, \beta)$-approximation algorithm if it runs in polynomial time and produces a solution of size at most $\alpha \cdot OPT$ while relaxing the constraint by a factor of $\beta$. 

\begin{figure}[t]
	\centering
	\begin{subfigure}[t]{0.48 \linewidth}
		\centering
		\includegraphics[page=1, width=0.9\textwidth]{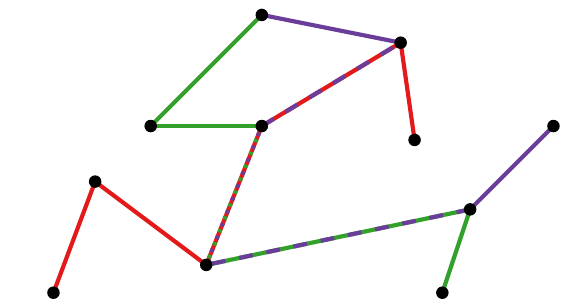}
		\caption{initial polyline bundle}
		\label{fig:exampleBiCritApprox-1}
	\end{subfigure}
	\hfill
	\begin{subfigure}[t]{0.48 \linewidth}
		\centering
		\includegraphics[page=2, width=0.9\textwidth]{exampleBiCritApprox}
		\caption{assigning a direction to each polyline}
		\label{fig:exampleBiCritApprox-2}
	\end{subfigure}
	
	\medskip
			
	\begin{subfigure}[t]{0.48 \linewidth}
		\centering
		\includegraphics[page=3, width=0.98\textwidth]{exampleBiCritApprox}
		\caption{all maximal stars}
		\label{fig:exampleBiCritApprox-3}
	\end{subfigure}
	\hfill
	\begin{subfigure}[t]{0.48 \linewidth}
		\centering
		\includegraphics[page=4, width=0.98\textwidth]{exampleBiCritApprox}
		\caption{greedy star cover of maximal stars}
		\label{fig:exampleBiCritApprox-4}
	\end{subfigure}
	
	\medskip
	
	\begin{subfigure}[t]{0.48 \linewidth}
		\centering
		\includegraphics[page=5, width=0.98\textwidth]{exampleBiCritApprox}
		\caption{retaining only bends of $B_\textrm{central} \cup B_\textrm{first}$}
		\label{fig:exampleBiCritApprox-5}
	\end{subfigure}
	\hfill
	\begin{subfigure}[t]{0.48 \linewidth}
		\centering
		\includegraphics[page=6, width=0.98\textwidth]{exampleBiCritApprox}
		\caption{resulting simplified polyline bundle}
		\label{fig:exampleBiCritApprox-6}
	\end{subfigure}
	
	\caption{Example of our bi-criteria $(\Oh(\log (\ell + n)), 2)$-approximation algorithm for \pbs.}
	\label{fig:exampleBiCritApprox}
\end{figure}

In our particular problem \pbs, we relax the error bound $\delta$.
In Section~\ref{sec:reductionFromMIDS}, we have shown that there is no bi-criteria $(n^{\frac{1}{3} - \varepsilon}, 1)$-approximation algorithm for \pbs for any $\varepsilon > 0$ unless \mbox{\p = \np}.
This strong inapproximability comes from the high sensitivity towards choices of keeping or discarding single bends, which is modulated by the given value of $\delta$.
By making a bad choice we cannot take (helpful) shortcuts that
have a distance just a little greater than the given distance
threshold~$\delta$ to the original sub-polyline. This can be overcome by relaxing the constraint slightly. In particular, we  show that allowing a constraint violation by a factor of $\beta=2$, we can design an efficient algorithm with an approximation guarantee of $ \alpha \in \Oh(\log(\ell + n ))$.
For an overview of our algorithm see Fig.~\ref{fig:exampleBiCritApprox}.

The key building block of our algorithm is a connection between \pbs and a certain geometric set cover problem, which we call \emph{star cover problem}. The star cover problem models the aspect of shortcutting polylines by few bend points but does not take into account consistency. We argue, however, that approximate solutions to the star cover problem can be post-processed to form consistent \pbs solutions by slightly violating the error threshold $\delta$.

\paragraph*{Star Cover Problem}

\begin{figure}[t]
	\centering
	\begin{minipage}[b]{0.48 \textwidth}
		\centering
		\includegraphics[scale=1.05]{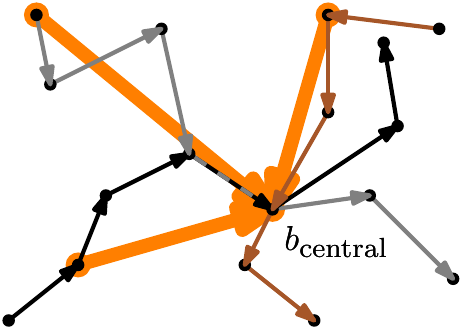}
		\caption{Example of a \emph{star} (in orange) around a bend $b_\textrm{central}$, which lies on three polylines. Each polyline was assigned an arbitrary direction indicated by arrow heads.}
		\label{fig:exampleStar}
	\end{minipage}
	\hfill
	\begin{minipage}[b]{0.48\textwidth}
		\centering
		\includegraphics[scale=1.05]{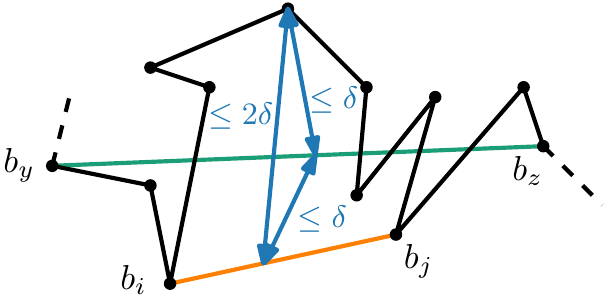}
		\caption{Example of the maximum Fr\'echet distance between a line segment $(b_i, b_j)$ and its corresponding sub-polyline if there is a valid shortcut $(b_y, b_z)$ going over $b_i$ and $b_j$.}
		\label{fig:lemmaDistanceSubsequence}
	\end{minipage}
\end{figure}

Next, we introduce the \emph{star cover} problem, which is a special type of the set cover problem defined over instances of \pbs.
Informally spoken, a star is a bend together with some incident shortcut segments. These shortcut segments span sets of original segments of the polylines. To this end, we first direct each polyline $L \in \mathcal{L}$ in a given \pbs instance $(B, \mathcal{L}, \delta)$ arbitrarily but ensuring that all (shortcut) segments of~$L$ are oriented in the same direction. Then a star consists of a set of incoming shortcuts of some bend; see Fig.~\ref{fig:exampleStar} for an example.
\begin{definition}[Star]
	A~\emph{star} is the combination of a bend $b_\textrm{central} \in B$ and, for each polyline $L \in \mathcal{L}$
	that contains~$b_\textrm{central}$, one or zero incoming shortcut segments (according to~$\delta$).
\end{definition}
We say a star $s$ \emph{covers} a segment--polyline pair~$(e, L)$, if $s$ contains for $L$ a shortcut  $(b_\textrm{outer}, b_\textrm{central})$ and $e$ lies on $L$ between $ b_\textrm{outer}$ and  $b_\textrm{central}$.
Our goal is to find a small set of stars that cover all segment--polyline pairs. We denote the set of all segment--polyline pairs in the input by $\mathcal{U}$ and the subset of pairs covered by a particular star $s$ by $\mathcal{U}_s$. Then the \emph{star cover problem} is defined as follows.

\begin{definition}[Star Cover]
	A \emph{star cover} $C$ is a set of stars, such that $\bigcup_{s \in C} \mathcal{U}_s = \mathcal{U}$, i.e. all segment--polyline pairs are covered. The \emph{star cover problem} (abbreviated by \starcover) asks for a minimum size star cover.
\end{definition}

\paragraph*{Relationship between Instances of Polyline Bundle Simplification and Star Cover}
Next, we investigate the relationship between an instance of \starcover and its corresponding instance of \pbs.
We argue that every (optimal) solution for \pbs can be decomposed into a star cover. Hence an optimal \starcover yields a lower bound for an optimal \pbs solution.

\begin{lemma}
	\label{lem:transformingInstancesOfPBStoSC}
	The size~$OPT_\starcover$ of an optimal solution of any instance of \starcover obtained from an instance~$(B, \mathcal{L}, \delta)$ of \pbs is bounded by~$OPT_\starcover \leq OPT_\pbs$, where $OPT_\pbs$ is the size of an optimal solution of~$(B, \mathcal{L}, \delta)$.
\end{lemma}

\begin{myproof}
	Consider an optimal solution~$B^*$ of~$(B, \mathcal{L}, \delta)$.
	From the simplified polyline bundle induced by~$B^*$, we can get a star cover for any instance of \starcover obtained from~$(B, \mathcal{L}, \delta)$
	by iteratively adding a star in the following way until there are only isolated bends.
	Get a star~$s$ by taking any connected bend~$b_\textrm{central} \subseteq B^*$ as a central bend and the bends that precede~$b_\textrm{central}$ on each of the simplified polylines as its outer bends.
	Remove the segment--polyline pairs covered by~$s$ from our simplified polyline bundle.
	Repeat this until there are no more segment--polyline pairs.
	The obtained star cover has at most $|B^*|$~stars and at least as many stars as a minimum star cover.
	So, $OPT_\starcover \leq OPT_\pbs$.
\end{myproof}

\paragraph*{Approximation for the Star Cover Problem}
We can compute an approximate solution for \starcover by employing the classical greedy algorithm~\cite{Johnson1974} for set cover,
which iteratively selects the set with the most uncovered elements until all elements are covered.
However, if applied naively, the running time would be exponential in the size of the \pbs instance as the number of stars might be in the order of $n \cdot 2^\ell$. We observe, however, that it suffices to consider only maximal stars (containing on each polyline incident to the central bend the incoming shortcut that covers the largest number of segments). As there are only $n$ maximal stars, this guarantees polynomial running time. 

\begin{lemma}
	\label{lem:star-cover}
	We can compute an $\Oh(\log (t + w))$-approximation for an instance of \starcover obtained from an instance~$(B, \mathcal{L}, \delta)$ in time~$\Oh(\ell n^3)$,
	where $t$ is the maximum number of polylines any bend point occurs in
	and $w$ is the maximum number of segments any valid shortcut (according to~$\delta$) can skip.
\end{lemma}

\begin{myproof}
	There is a polynomial time greedy algorithm that yields
	an $\Oh(\log m)$ approximation for the set cover problem,
	where $m$ is the size of the largest set in the given collection of
	subsets of the universe~\cite{Johnson1974}.
	The greedy algorithm works as follows.
	While there are uncovered elements from the universe,
	add the set with the largest number of uncovered elements to the set cover.
	In an instance of \starcover, this $m$ is the maximum number of segment--polyline pairs~$\max_{\textrm{star } s}|\mathcal{U}_s|$ a single star can cover.
	If the central bend point of a star lies in at most~$t$ polylines, the star contains at most $t$ shortcut segments, and each of which covers at most $w$ segments, hence we have $m = t w$.
	Observe that $\Oh(\log (t w)) = \Oh(\log (t + w))$.
	
	Having settled the $\Oh(\log (t + w))$ approximation ratio,
	it remains to prove the polynomial running time.
	Using the algorithm by Imai and Iri~\cite{Imai1988} independently for each polyline,
	we can find all (maximal) shortcuts for every bend on every polyline in time~$\Oh(\ell n^3)$.
	Combining these shortcuts at every bend gives us all $n$ maximal stars in time~$\Oh(\ell n)$.
	For each star, we also save the number of segment--polyline pairs it covers 
	and, to each segment--polyline pair, we link all stars it appears in.
	Both can be done in time~$\Oh(\ell n^2)$.
	As long as there are uncovered segments,
	we find the star with the most uncovered segments and then update
	the number of uncovered segments for the other stars.
	This can be done in $\Oh(\ell n^2)$ time in total as well.
\end{myproof}

\paragraph*{Relationship between Star Covers and Solutions of Polyline Bundle Simplification}

While a solution for \pbs can be directly converted into a star cover as argued above, the converse is more intricate.
The shortcuts contained in the selected stars may be overlapping or nested along a polyline, that is, bends skipped by one shortcut may be end points of another shortcut in the set.
Moreover, shared parts of different polylines may be shortcut differently.
Therefore consistency is not guaranteed.
We explain how to derive from a star cover solution a solution for its corresponding instance of~\pbs.
Some of the shortcuts of the \starcover solution are replaced by shorter shortcuts in order to integrate some intermediate point to the \pbs solution.
Lemma~\ref{lem:innerBendsMaxDist2} states that those newly introduced shortcuts can be at most $2\delta$ away from the original polyline.
The situation described there is depicted in Fig.~\ref{fig:lemmaDistanceSubsequence}.
It follows immediately from a lemma by Agarwal et al.~(\cite{Agarwal2005}, Lemma~3.3).

\begin{lemma}
	\label{lem:innerBendsMaxDist2}
	Given a polyline~$L = (b_1, b_2, \dots, b_{|L|})$ and a distance threshold~$\delta$.
	If there are $y, z \in \mathbb{N}$ with $1 \leq y < z \leq |L|$ and $d_\textnormal{Fr\'echet}((b_y, b_z), L[b_y, \dots, b_z]) \leq \delta$ (i.e., segment $(b_y, b_z)$ is a valid shortcut),
	then for any $i, j \in \mathbb{N}$ with $y \leq i < j \leq z$,
	$d_\textnormal{Fr\'echet}((b_i, b_j), L[b_i, \dots, b_j]) \leq 2 \delta$.
\end{lemma}

Equipped with this lemma, we now discuss the actual transformation from a \starcover solution to a \pbs solution. The idea is to keep, beside the starting points of all polylines, only the central bend points of the selected stars while dropping their leaves. This is closely tied with the fact that we minimize the number of stars while ignoring their degree in the algorithm. The main insight here is that the shortcuts induced by this augmented point set still have a small distance to the original~polylines.

\begin{lemma}
	\label{lem:scSolution2pbsSolution}
	Let $C$ be a star cover for an instance of \starcover obtained from an instance~$(B, \mathcal{L}, \delta)$ of \pbs.
	If $C$ is an $\alpha$-approximation for its instance of~\starcover, a bi-criteria $(\alpha + 1, 2)$-approximation for $(B, \mathcal{L}, \delta)$ can be computed in time~$\Oh(n)$ from~$C$.
\end{lemma}

\begin{myproof}
	Let $B_\textrm{central}$ be the set of central bends of the stars in~$C$ and let $B_\textrm{first}$ be the set of first bends of all polylines from~$\mathcal{L}$.
	We return $B_\textrm{central} \cup B_\textrm{first}$ as the bi-criteria approximate solution.
	Clearly, we can construct this set in time~$\Oh(n)$.
	According to Lemma~\ref{lem:transformingInstancesOfPBStoSC}, $OPT_\starcover \leq OPT_\pbs$, where $OPT_\pbs$ is the size of the optimal solution of $(B, \mathcal{L}, \delta)$ and $OPT_\starcover$ is the size of the optimal solution of the instance of~\starcover where~$C$ is an approximation for.
	We conclude
	\begin{align}
	|B_\textrm{central} \cup B_\textrm{first}| \leq \alpha OPT_\starcover + OPT_\pbs \leq (\alpha + 1) OPT_\pbs \, .
	\end{align}
	Let $\mathcal{L}'$ be the polyline bundle induced by $B_\textrm{central} \cup B_\textrm{first}$.
	It remains to prove that the Fr\'echet distance between each induced segment of each polyline in $\mathcal{L}'$ and its corresponding sub-polyline in $\mathcal{L}$ is at most~$2 \delta$.
	Consider any segment~$(b_i, b_j)$ of any polyline $L' \in \mathcal{L}'$ corresponding to a polyline $L \in \mathcal{L}$ such that $b_i$ precedes~$b_j$ in~$L$.
	There is a star~$s$ in~$C$ that covers all segments of $L[b_i, b_j]$.
	Clearly, all segments of $L[b_i, b_j]$ are covered by the stars of~$C$ and if there was no single star~$s$ covering all segments of $L[b_i, b_j]$, but multiple stars,
	there would be another central bend of a star between~$b_i$ and~$b_j$ on~$L$ and, in~$L'$, $(b_i, b_j)$ would not be a segment.
	The central bend~$b_\textrm{central}$ of~$s$ succeeds~$b_j$ or is equal to~$b_j$ as otherwise $s$ would not cover all of $L[b_i, b_j]$.
	Accordingly, the outer bend~$b_\textrm{outer}$ of~$s$ on~$L$ precedes~$b_i$ or is equal to~$b_i$ as otherwise $s$ would not cover all of $L[b_i, b_j]$.
	By the definition of a star, we know that $d_\textnormal{Fr\'echet}((b_\textrm{outer}, b_\textrm{central}), L[b_\textrm{outer}, b_\textrm{central}]) \leq \delta$.
	By Lemma~\ref{lem:innerBendsMaxDist2}, it follows that $d_\textnormal{Fr\'echet}((b_i, b_j), L[b_i, b_j]) \leq 2 \delta$.
\end{myproof}

\paragraph*{Bi-criteria Approximation for Polyline Bundle Simplification via Star Cover}
Using the previous lemmas, we obtain the main theorem of this section. It is reasonable to assume that the number $\ell$ of polylines is polynomial in $n$ in practically relevant settings. Hence, we essentially obtain an exponential improvement over the complexity-theoretic lower bound $n^{\frac{1}{3}-\varepsilon}$ if we allow the slight violation of the error bound.

\begin{theorem}
	\label{thm:bi-crit-algo-pbs}
	There is a bi-criteria $(\Oh(\log (\ell + n)), 2)$-approximation algorithm for \pbs running in time~$\Oh(\ell n^3)$,
	where $\ell$ is the number of polylines and $n$ is the number of bend points in the polyline bundle.
\end{theorem}

\begin{myproof}
	We describe a (kind of) approximation-preserving reduction from \pbs to \starcover,
	which can be realized as a bi-criteria approximation algorithm.
	Its steps are depicted in Fig.~\ref{fig:exampleBiCritApprox}.
	Given an instance $(B, \mathcal{L}, \delta)$ of \pbs, where we let the size of the optimal solution be $OPT_\pbs$, we assign an arbitrary direction to each $L \in \mathcal{L}$.
	This yields our corresponding instance of \starcover.
	For this corresponding instance of \starcover, compute an $\Oh(\log(t+w))$ approximation star cover~$C$.
	We can do this in time~$\Oh(\ell n^3)$ according to Lemma~\ref{lem:star-cover}.
	According to Lemma~\ref{lem:scSolution2pbsSolution}, we can compute a bi-criteria $(\Oh(\log (t + w)), 2)$-approximation for $(B, \mathcal{L}, \delta)$ from~$C$ in $\Oh(n)$ time.
	Since $t \leq \ell$ and $w \leq n$, this is also a bi-criteria $(\Oh(\log (\ell + n)), 2)$-approximation.
\end{myproof}

\section{Fixed-Parameter Tractability}
\label{sec:exact}
A brute force approach is checking for every subset of the bend set~$B$ in time $\Oh(\ell \cdot n)$ whether it is a valid simplification and accepting the one with the smallest number of bends or segments.
Consequently, the runtime of this approach is $\Oh(2^n \cdot \ell \cdot n)$.
When considering fixed-parameter tractability, investigating parameters of the input is a natural choice.
According to Corollary~\ref{cor:pbsNotFPTinL}, \pbs is not fixed-parameter tractable (FPT) in the number of polylines $\ell$.
However, \pbs is FPT in the number of shared bends, i.e., bends contained in more than one polyline. We denote the set of those bends by~$B_\textrm{shared}$ and we let $k:=|B_\textrm{shared}|$.

\begin{restatable}[]{theorem}{theoremFPT}
	\label{thm:fpt-in-V-shared}
	\pbs is FPT in the number of shared bends~$k$.
	There is an algorithm solving \pbs in time~$\Oh(2^{k} \cdot \ell \cdot n^2 + \ell n^3)$. 
\end{restatable}


\begin{myproof}
	We describe an algorithm that solves \pbs in time $\Oh(2^{k} \cdot \ell \cdot n^3)$.
	Given an instance~$(B, \mathcal{L}, \delta)$ of \pbs, the first step is to compute, for each $L \in \mathcal{L}$, its shortcut graph~$G_L$ using the algorithm by Imai and Iri~\cite{Chan1996}.
	This can be done in time~$\Oh(\ell \cdot n^3)$.
	For a polyline~$L$ and a distance threshold~$\delta$,
	the \emph{shortcut graph} is the directed graph that has the bends of~$L$ as its vertices and has an edge from~$u$ to $v$ if $d_\textnormal{Fr\'echet}((u,v), L[u,\cdots,v]) \leq \delta$, this is, if there is a shortcut from~$u$ to~$v$ in~$L$.
	Given the shortcut graph~$G_L$ of~$L$, the vertices of a shortest path in~$G_L$ from the first bend of~$L$ to the last bend of~$L$ define an optimal simplification of~$L$.
	
	The second step is to iterate over all subsets $B' \subseteq B_\textrm{shared}$ and check if~$B'$ is part of an optimal solution.
	Before the first iteration, we initialize a variable $n_\textrm{min} = \infty$ and we will save the current best solution by~$\mathcal{S}_\textrm{min}$.
	Then, in each iteration, we temporarily remove from all shortcut graphs $G_L$ all vertices~$B_\textrm{not-contained} = B_\textrm{shared} - B'$ and all edges that correspond to a shortcut skipping a bend in~$B'$.
	Clearly, removing $B_\textrm{not-contained}$ can be performed in $\Oh(n^2)$ time for each~$G_L$.
	For the removal of the edges in~$G_L$, note that we can sort the list of bends~$B_\textrm{not-contained}$ and the list of all edges (defined by their endpoints) alphanumerically by the occurrence of the bends within the polyline~$L$.
	If we traverse both lists simultaneously in ascending order, we remove an edge if and only if its endpoint-bends come before and after the currently considered bend from~$B_\textrm{not-contained}$.
	Therefore, the removal operations can be performed in~$\Oh(n^2)$ time per~$G_L$.
	
	If some shortcut graph becomes disconnected by these removal operations, we continue with the next iteration.
	Otherwise, we take the bends of a shortest path from the first to the last bend in each reduced version of $G_L$.
	Together they define a simplification~$\mathcal{S}$ of our \pbs instance.
	If the number~$n_\mathcal{S}$ of bends in~$\mathcal{S}$ is less than~$n_\textrm{min}$, we set $n_\textrm{min} = n_\mathcal{S}$ and $\mathcal{S}_\textrm{min} = \mathcal{S}$.
	After the iteration process, we return~$\mathcal{S}_\textrm{min}$.
	Since we have $2^k$ subsets of $B_\textrm{shared}$ and each iteration can be performed in~$\Oh(\ell \cdot n^2)$ time, the running time of the algorithm is in~$\Oh(2^{k} \cdot \ell \cdot n^2 + \ell n^3)$.
	
	It remains to prove that $\mathcal{S}_\textrm{min}$ is in the end an optimal solution of our input instance of \pbs.
	First note that our algorithm always returns some polyline simplification because for $B' = B_\textrm{shared}$, we do not get a disconnected~$G_L$ after the removal operations.
	
	The returned solution is valid because the shared bends of~$B'$ are taken in all simplified polylines (they cannot be skipped) and the other shared bends are skipped in all simplified polylines.
	Our algorithm finds the minimum size solution because in one iteration it considers $B' = B^* \cap B_\textrm{shared}$, where $B^*$ is the set of retained bends of an optimal solution.
	Moreover, an optimal solution cannot have fewer bends occurring in only one polyline~$L$ than our algorithm since this would imply a shorter shortest path within the reduced version of~$G_L$.
\end{myproof}

\section{Conclusion and Outlook}
\label{sec:conclusion-and-outlook}

\begin{figure}[t]
	\centering
	\begin{subfigure}[t]{0.32 \linewidth}
		\centering
		\includegraphics[page=1, width=1.0 \linewidth]{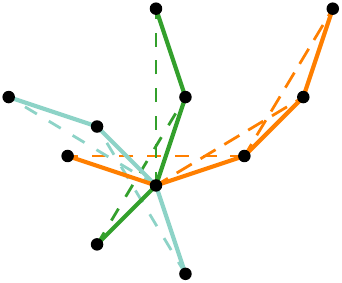}
		\caption{initial bundle with shortcuts}
		\label{fig:exampleMinVEdifferent-base}
	\end{subfigure}
	\hfill
	\begin{subfigure}[t]{0.32 \linewidth}
		\centering
		\includegraphics[page=3, width=1.0 \linewidth]{exampleMinVEdifferent}
		\caption{optimal for \minBends \nolinenumbers}
		\label{fig:exampleMinVEdifferent-minE}
	\end{subfigure}
	\hfill
	\begin{subfigure}[t]{0.32 \linewidth}
		\centering
		\includegraphics[page=2, width=1.0 \linewidth]{exampleMinVEdifferent}
		\caption{optimal for \minSegments \nolinenumbers}
		\label{fig:exampleMinVEdifferent-minV}
	\end{subfigure}
	
	\caption{Example of three polylines, where the goals \minSegments and \minBends differ.}
	\label{fig:exampleMinVEdifferent}
\end{figure}

We have generalized the well-known problem of polyline simplification from a single polyline to polyline bundles.
Although in the case of one polyline, efficient algorithms have long been known,
it turned out that simplifying two or more polylines is a problem that is indeed hard to approximate within a factor of $n^{\frac{1}{3} - \varepsilon}$ for any $\varepsilon > 0$.
However, if we relax the constraint on the maximum Fr\'echet distance between original and simplified polyline by a factor of~2, we can overcome this strong inapproximability bound.
Moreover, we can find an optimal simplification quickly if we have only a small number of shared bends since the problem of polyline bundle simplification is fixed-parameter tractable (FPT) in this parameter.

Based on our results, there are many possible directions for future~research.
\begin{itemize}
	\item Our current bi-criteria approximation guarantee is logarithmic in the number of polylines~$\ell$ plus the number of bend points $n$. In most practical application, $\ell$ is smaller than $n$ or at most polynomial in $n$. From a theoretical perspective, however, it might be interesting to  get rid off the dependency on $\ell$ in the bi-criteria approximation in order to get improvements for the case where $\ell$ is significantly larger than $n$.
	\item As a distance measure, we employed the Fr\'echet distance, which we consider to be more natural and intuitive than the Hausdorff distance when comparing polylines.
	However, the Hausdorff distance is sometimes used in classical polyline simplification as well.
	Our hardness results also apply to the Hausdorff distance,
	but our bi-criteria approximation algorithm fails since Lemma~\ref{lem:innerBendsMaxDist2} is not true for the Hausdorff distance.
	One might consider \pbs using the Hausdorff distance or other (even non-segment-wise) distance measurements.
	
	\item In our generalization to bundles of polylines, we aim for a minimizing the number of retained bends (\minBends).
	However, minimizing the number of retained segments (\minSegments) is an alternative goal, which also generalizes the classical minimization problem for a single polyline.
	Optimal simplifications for both goals may differ; see~Fig.~\ref{fig:exampleMinVEdifferent}.
	Our hardness and FPT results also apply for the goal~\minSegments.	However, it is not clear how to obtain a similar result for the bi-criteria approximability.
	
	\item For practical purposes, the  scalability of the proposed bi-criteria approximation algorithm, the FPT algorithm, and possibly new heuristics should be investigated on real-world~data.
\end{itemize}



\bibliography{mybibliography.bib}

\clearpage
\appendix

{\Large\bf\noindent Appendix}

\section{Omitted Content of Section~\ref{sec:reductionFromMIDS}}
\label{app:reductionFromMIDS}

\begin{figure}[t]
	\captionsetup[subfigure]{justification=centering}
	\centering
	\begin{subfigure}[b]{0.26 \linewidth}
		\centering
		\includegraphics[page=2,scale=1.1,trim=235 147 529 0,clip]{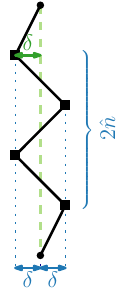}
		\caption{Vertex gadget.}
		\label{fig:appendix-MIDS-reduction-vertex-gadget}
	\end{subfigure}
	\hfill
	\begin{subfigure}[b]{0.7 \linewidth}
		\centering
		\includegraphics[page=4,scale=1.24,trim=226 177 372 34,clip]{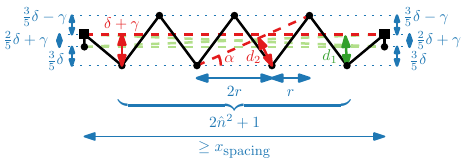}
		\caption{Edge gadget.}
		\label{fig:appendix-MIDS-reduction-edge-gadget}
	\end{subfigure}
	
	\medskip
	
	\begin{subfigure}[b]{1 \linewidth}
		\centering
		\includegraphics[page=12,scale=1.4,trim=264 0 276 205,clip]{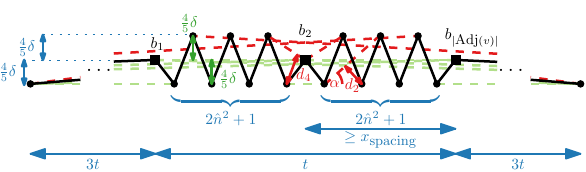}
		\caption{Neighborhood gadget.}
		\label{fig:appendix-MIDS-reduction-neighborhood-gadget}
	\end{subfigure}
	\caption{
		Some additional details to Fig.~\ref{fig:MIDS-reduction}.
	}
	\label{fig:appendix-MIDS-reduction}
\end{figure}

It remains to show the correctness of Claim~\ref{clm:claimEdgegadget} and Claim~\ref{clm:claimNeighborhoodgadget}, 
which we use in our reduction from \mids to \pbs.
Our gadgets are depicted in Fig.~\ref{fig:MIDS-reduction}.
For convenience, we provide by Fig.~\ref{fig:appendix-MIDS-reduction} a copy of them with some additional details, to which we will refer in this appendix.
For example~$r$ is the x-distance between two consecutive (inner) vertices in an edge and a neighborhood gadget (if a gadget is rotated, the distance is measured along the corresponding rotated axis).
We know that $r \ge x_\textrm{spacing} / (2\hat{n}^2+2)$.

\begingroup
\def\thetheorem{\ref{clm:claimEdgegadget}}
\claimEdgegadget*
\addtocounter{theorem}{-1}
\endgroup

In~(i), both of the shared bends, these are the second and the second last, are skipped and we can take the ``long'' shortcut from the first to the last bend because the line segment between them is horizontal and has y-distance $3/5 \delta$ or $2/5 \delta + \gamma$ or $\delta$ to all inner bends.
In~(ii), the most critical part is the distance~$d_1$ between the third last bend and the straight-line segment from the first to the second last bend (see Figure~\ref{fig:MIDS-reduction-edge-gadget}).
It is
\begin{equation}
	d_1 \leq \frac{3}{5} \delta + \left( \frac{2}{5} \delta + \gamma \right) - \frac{\frac{2}{5} \delta + \gamma}{2\hat{n}^2+2} \leq \delta + \gamma - \frac{\frac{2}{5} \cdot \frac{10 \hat{n}^2 + 5}{2} \gamma + \gamma}{2 \hat{n}^2+2} = \delta \, .
\end{equation}
Observe that (iii) is the same as (ii) but mirrored.
If neither the second nor the second last bend is skipped, i.e., if $u$ and $v$ are in the set $V^*$, then we cannot cut short anything in this gadget.
Clearly, we cannot take a ``long'' shortcut from the second to the second last bend because the lower row of inner bends has distance $\delta + \gamma$ from the potential shortcut segment.
Moreover, we cannot take a ``short'' shortcut from a bend of the lower row to a bend of the upper row or the other way around.
If we would aim to skip two inner bends, the distance~$d_2$ (see Figure~\ref{fig:MIDS-reduction-edge-gadget}) from an inner bend to the shortcut segment would have to be at most $\delta$.
However, it is
\begin{equation}
\label{eq:d2-step1}
	d_2 = \sin \alpha \cdot 2 r = \sin \arctan \frac{\frac{8}{5} \delta}{3 r} \cdot 2 r = \frac{\frac{8 \delta}{15 r}}{\sqrt{\left(\frac{8 \delta}{15 r}\right)^2 + 1}} \cdot 2 r = \frac{16 \delta r}{\sqrt{(8 \delta)^2 + (15 r)^2}} \, ,
\end{equation}
where
\begin{equation}
\label{eq:d2-step2}
	r \geq \frac{x_\textrm{spacing}}{2\hat{n}^2 + 2} \geq \frac{(2\hat{n}^2+2) 3\delta}{2\hat{n}^2+2} = 3 \delta \, ,
\end{equation}
and hence,
\begin{equation}
\label{eq:d2-step3}
	d_2 \geq \frac{48 \delta^2}{\sqrt{64 \delta^2 + 2025 \delta^2}} = \frac{48}{\sqrt{2089}} \delta = 1.0502\dots \delta \, .
\end{equation}

\begin{figure}
	\centering
	\includegraphics[page=13,scale=1.0,trim=180 74 290 90,clip]{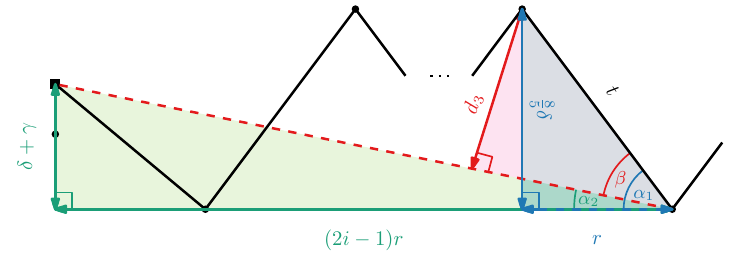}
	\caption{The potential shortcut segment in an edge gadget from the second bend to an inner bend (the $(2i+1)$-th bend) is dashed in red.
		However, $d_3 > \delta$ makes it no valid shortcut segment.}
	\label{fig:MIDS-reduction-edge-gadget-special-shortcut}
\end{figure}

Observe that this becomes even greater if we aim for skipping four or more bends or if we start or end at one of the two shared bends.
To make this clearer, we explicitly consider the latter case where a potential shortcut would start at the second bend and end at the~$(2i + 1)$-th bend.
This situation is depicted in Fig.~\ref{fig:MIDS-reduction-edge-gadget-special-shortcut}.
If it was a valid shortcut, $d_3$ would be less than or equal to~$\delta$.
Since~$d_3$ is inside a rectangular triangle, its length is
\begin{equation}
	d_3 = t \cdot \sin \beta \, ,
\end{equation}
where $t$ is inside another rectangular triangle with legs of length~$r$ and $8/5 \delta$, so
\begin{equation}
	t = \sqrt{r^2 + \left( \frac{8}{5} \delta \right)^2} \, .
\end{equation}
We can determine $\beta$ via the angles~$\alpha_1$ and $\alpha_2$ as
\begin{equation}
	\beta = \alpha_1 - \alpha_2 = \arctan \frac{\frac{8}{5} \delta}{r} - \arctan \frac{\delta + \gamma}{(2i - 1) r} \, .
\end{equation}
In the $\arctan$-functions, all parameters are positive, so they live in the range $[0, \pi/2)$.
Hence, $\beta$ lives in the range $(-\pi / 2, \pi / 2)$.
In this range, the $\sin$-function is monotonously increasing.
Therefore, to give a lower bound on $d_3$, we can use a lower bound on $\sin \beta$ by specifying a lower bound on $\beta$.
Since $i \ge 2$, $\gamma < \delta/(5 \hat{n}^2)$ and $\hat{n} \ge 1$, we state that
\begin{equation}
	\beta = \alpha_1 - \alpha_2 > \arctan \frac{\frac{8}{5} \delta}{r} - \arctan \frac{\frac{6}{5} \delta}{3 r} = \arctan \frac{8}{5r'} - \arctan \frac{2}{5 r'} \, ,
\end{equation}
where $r' = r/\delta$.
A lower bound on $t$ is
\begin{equation}
	t = \sqrt{\left(r' \delta\right)^2 + \left( \frac{8}{5} \delta \right)^2} > r' \delta \, .
\end{equation}
So, we can get a lower bound on $d_3$ by
\begin{equation}
	d_3 = t \cdot \sin \beta > \underbrace{r' \sin\left( \arctan \frac{8}{5r'} - \arctan \frac{2}{5 r'} \right) }_{c(r')} \cdot \hspace{3pt} \delta \, .
\end{equation}
To prove that $d_3$ is always greater than $\delta$, it suffices to show that the prefactor~$c(r')$ is equal to or greater than~$1$ for all possible values of~$r'$.
We reformulate $c(r')$ using well-known trigonometric identities:
\begin{align}
	c(r') &= r' \sin\left( \arctan \frac{8}{5r'} - \arctan \frac{2}{5 r'} \right) \nonumber \\
	&= \frac{6}{\sqrt{25 + \frac{68}{r'^2} + \frac{256}{25 r'^4}}}
	\label{eq:c(r')}
\end{align}
For~$r' = 3$, this is $c(r') = 1.0495\dots$ and, from Equation~(\ref{eq:c(r')}), it is easy to see that $c(r')$ is even greater for $r' > 3$.
Thus, we conclude that $d_3 > \delta$ always holds.


It remains to consider potential shortcuts starting or ending at the first or the last bend.
Clearly, skipping only the second or second last bend is always possible.
Skipping the second and the third bend or skipping the second last and the third last bend may sometimes be possible depending on how much the edge gadget is stretched horizontally.
However, according to the previous analysis, skipping more bends is not possible since the distance between the potential shortcut segment and the bend before the end point of the potential shortcut is at least~$d_3$.

\begingroup
\def\thetheorem{\ref{clm:claimNeighborhoodgadget}}
\claimNeighborhoodgadget*
\addtocounter{theorem}{-1}
\endgroup

\begin{figure}
	\centering
	\includegraphics[page=14,scale=1.0,trim=180 58 290 90,clip]{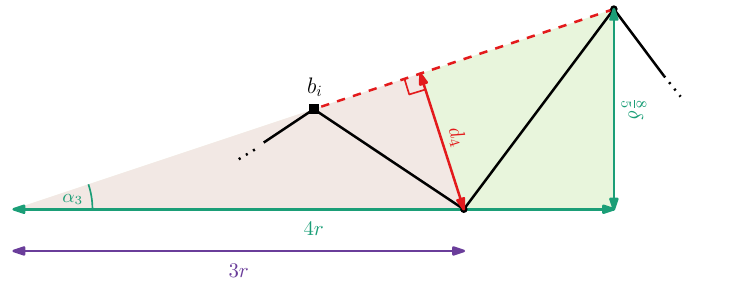}
	\caption{The potential shortcut segment in a neighborhood gadget form a bend~$b_i$ to an inner bend is dashed in red.
	However, $d_4 > \delta$ makes it no valid shortcut segment.}
	\label{fig:MIDS-reduction-neighborhood-gadget-special-shortcut}
\end{figure}

Clearly, the shortcuts~(i) for skipping any $b_i$ (or exactly one neighbor of $b_i$) are valid and there is no shortcut from the first to the last bend since the potential shortcut segment has distance $8/5 \delta$ to the upper row of bends.
In~(ii), there clearly is a shortcut if we start at any $b_i$ and end at any~$b_j$.
If we start at some~$b_i$ and end at the last bend, observe that, in the most extreme case, the segment from $b_1$ to the last bend has a y-distance to the upper row of
\begin{equation}
\frac{4}{5} \delta + \frac{t}{4 t} \cdot \frac{4}{5} \delta = \delta
\end{equation}
when it passes $b_{|\textrm{Adj}(v)|}$ in x-dimension.
Thus, this shortcut is valid and the same holds for the shortcuts from the first bend to some $b_j$.

It remains to argue that there are no more shortcuts.
A shortcut starting and ending at a bend on the upper or lower row is not possible because it would either be a horizontal segment, which has distance $8/5 \delta$ to the other row, or the distance to some bend in between would be at least~$d_2$, which we have shown to be greater than~$\delta$ in Equations~(\ref{eq:d2-step1})--(\ref{eq:d2-step3}).
It is easy to see that there is no shortcut starting at the first bend and ending at some inner bend of the upper or lower row.
The same holds true for shortcuts starting at some inner bend of the upper or lower row and ending at the last bend.

Moreover, a shortcut segment starting (ending) at some $b_i$ for $i \in \{1, \dots, |\textrm{Adj}(v)|\}$ and skipping one bend would have a distance of~$d_4$ to this bend as depicted in Fig.~\ref{fig:MIDS-reduction-neighborhood-gadget-special-shortcut}.
Since~$d_4$ is inside a rectangular triangle, we can determine~$d_4$ by
\begin{equation}
d_4 = 3 r \cdot \sin \alpha_3 \, ,
\end{equation}
where $\alpha_3$ is in another rectangular triangle and thus can be determined by
\begin{equation}
\alpha_3 = \arctan \frac{\frac{8}{5} \delta}{4r} = \arctan \frac{2}{5 r'} \, .
\end{equation}
Putting them together, we get
\begin{align}
d_4 &= 3 r' \delta \cdot \sin \arctan \frac{2}{5 r'} = 3 r' \delta \frac{\frac{2}{5 r'}}{\sqrt{1 - \frac{4}{25 r'^2}}} = \frac{6}{\sqrt{25 - \frac{4}{r'^2}}} \delta \, .
\end{align}
For $r' = 3$, this is $1.2108\dots \delta$ and again, for $r' > 3$, $d_4$ is even greater.

If we skip more than one inner bend, the distance to the last skipped bend becomes only greater.
Hence, we conclude that Claim~\ref{clm:claimNeighborhoodgadget} is correct.

\end{document}